\documentclass[aps,prl,twocolumn,superscriptaddress,showpacs]{revtex4}
\usepackage{graphicx}
\usepackage{amssymb}
\usepackage{amsmath}

\usepackage{pdfpages}

\begin{document}

\title{Gap symmetry of the heavy fermion superconductor CeCu$_2$Si$_2$ at ambient pressure}

\author{Yu Li}
\affiliation{Beijing National Laboratory for Condensed Matter Physics and
Institute of Physics, Chinese Academy of Sciences, Beijing 100190, China}
\affiliation{University of Chinese Academy of Sciences, Beijing 100049, China}
\author{Min Liu}
\affiliation{Beijing National Laboratory for Condensed Matter Physics and
Institute of Physics, Chinese Academy of Sciences, Beijing 100190, China}
\affiliation{College of Physical Science and Technology, Sichuan University, Chengdu 610065, China}
\author{Zhaoming Fu}
\affiliation{College of Physics and Material Science, Henan Normal University, Xinxiang 453007, China}
\author{Xiangrong Chen}
\affiliation{College of Physical Science and Technology, Sichuan University, Chengdu 610065, China}
\author{Fan Yang}
\affiliation{School of Physics, Beijing Institute of Technology, Beijing, 100081, China}
\author{Yi-feng Yang}
\email[]{yifeng@iphy.ac.cn}
\affiliation{Beijing National Laboratory for Condensed Matter Physics and
Institute of Physics, Chinese Academy of Sciences, Beijing 100190, China}
\affiliation{University of Chinese Academy of Sciences, Beijing 100049, China}
\affiliation{Collaborative Innovation Center of Quantum Matter, Beijing 100190, China}

\date{\today}

\begin{abstract}
Recent observations of two nodeless gaps in superconducting CeCu$_2$Si$_2$ have raised intensive debates on its exact gap symmetry, while a satisfactory theoretical basis is still lacking. Here we propose a phenomenological approach to calculate the superconducting gap functions, taking into consideration both the realistic Fermi surface topology and the intra- and interband quantum critical scatterings. Our calculations yield a nodeless $s^\pm$-wave solution in the presence of strong interband pairing interaction, in good agreement with experiments. This provides a possible basis for understanding the superconducting gap symmetry of CeCu$_2$Si$_2$ at ambient pressure and indicates the potential importance of multiple Fermi surfaces and interband pairing interaction in understanding heavy fermion superconductivity.
\end{abstract}

\maketitle

Recent experiments on the first heavy fermion superconductor CeCu$_2$Si$_2$ \cite{Steglich1979} have stimulated heated debates on the exact symmetry of its superconducting gap \cite{Kittaka2014}. The superconductivity of the so-called $S$-type CeCu$_2$Si$_2$ has long been believed to be of nodal $d$-wave with a transition temperature $T_c\approx 0.6\,$K at ambient pressure. Earlier nuclear magnetic resonance (NMR) measurements revealed a significant drop of the Knight shift below $T_c$ \cite{Ueda1987}, indicating the spin-singlet nature of its superconducting pairing. At intermediate temperatures below $T_c$, both the specific-heat coefficient $C/T$ and the NMR spin-lattice relaxation rate $1/T_1$ seem to exhibit power-law temperature dependence: $C/T\propto T$ \cite{Bredl1983,Arndt2011} and $1/T_1\propto T^3$ \cite{Kitaoka1986,Fujiwara2008}, suggesting the possible existence of line nodes. Neutron scattering experiments reported a broad spin resonance below $T_c$ \cite{Stockert2011}, which is a signature of sign change of the superconducting gap function on the Fermi surfaces \cite{Scalapino2012}. The upper critical field was also found to exhibit fourfold oscillation \cite{Vieyra2011}. These observations have led to the belief that superconductivity of CeCu$_2$Si$_2$ at ambient pressure should be of $d$-wave, which naturally arises if it emerges at the border of spin density wave antiferromagnetic order \cite{Yuan2003,Steglich2005} and is mediated by magnetic quantum critical fluctuations on a single nested heavy electron Fermi surface \cite{Eremin2008}. A propagation vector $\bold{Q}_\text{AF}=(0.215,0.215,0.53)$ has been observed in the neutron scattering experiment \cite{Stockert2004}, in agreement with renormalized band calculations \cite{Zwicknagl1993}.

This simple scenario has been lately questioned by a number of refined experiments. The specific-heat coefficient $C/T$ was measured again down to $60\,$mK on high-quality samples, revealing an exponential $T$-dependence at very low temperatures that could be fitted by two nodeless gaps with a ratio of about 2.5 \cite{Kittaka2014}. Scanning tunneling microscopy and spectroscopy (STM/STS) measurements down to 20 mK also showed clear evidence of two gaps \cite{Enayat2016}. Other measurements including angle-resolved specific heat \cite{Kittaka2016}, London penetration depth \cite{Pang2016,Takenaka2017}, as well as thermal conductivity \cite{Yamashita2017}, all pointed to similar conclusions and led to a variety of different proposals including $s^{\pm}$, $s^{++}$, and '$d+d$'-wave. Yet the exact gap symmetry remains undecided and a concrete theoretical basis is still lacking.

Overall, the above refined experiments highlight the importance of multiple Fermi surfaces. It is therefore crucial to go beyond the single-band scenario and take into account realistic band structures of CeCu$_2$Si$_2$. This was first considered in Ref.~\cite{Ikeda2015} by using the density functional theory (DFT)+$U$ together with the random phase approximation or second-order perturbation for the dynamical susceptibility. In the latter case, they could obtain a $s^{\pm}$-wave solution with loop nodes. While their results revealed the importance of a light hole Fermi surface in addition to the heavy electron one, the existence of loop nodes is inconsistent with experiments.

Here we resolve this remaining discrepancy by considering a phenomenological approach that combines the realistic multiple Fermi surfaces and a quantum critical form of the magnetic susceptibility \cite{Monthoux2007,Yang2014}. With varying strengths of the inter- and intraband interactions, we are able to extract the most important factors that govern the appearance of different gap symmetries. By solving the two-band Eliashberg equations, we find that a nodeless $s^{\pm}$-wave solution becomes dominant for a strong interband pairing interaction and the derived gap ratio is consistent with experimental estimates; while for weak interband interaction, we reproduce the usual $d_{x^2-y^2}$-wave gap on the heavy electron Fermi surface. In between, we find a nodal $s$-wave solution which is associated primarily with the hole Fermi surface. This explains the appearance of either $d_{x^2-y^2}$ or nodal $s$-wave solutions in previous theories and points to the critical role of an interband pairing interaction in determining the superconducting properties of CeCu$_2$Si$_2$. Since this is usually ignored in previous discussions, our understanding of superconductivity may need to be revisited for all realistic heavy fermion superconductors with complex Fermi surfaces and strong interband scattering.

\begin{figure}[t]
\centering\includegraphics[width=0.48\textwidth]{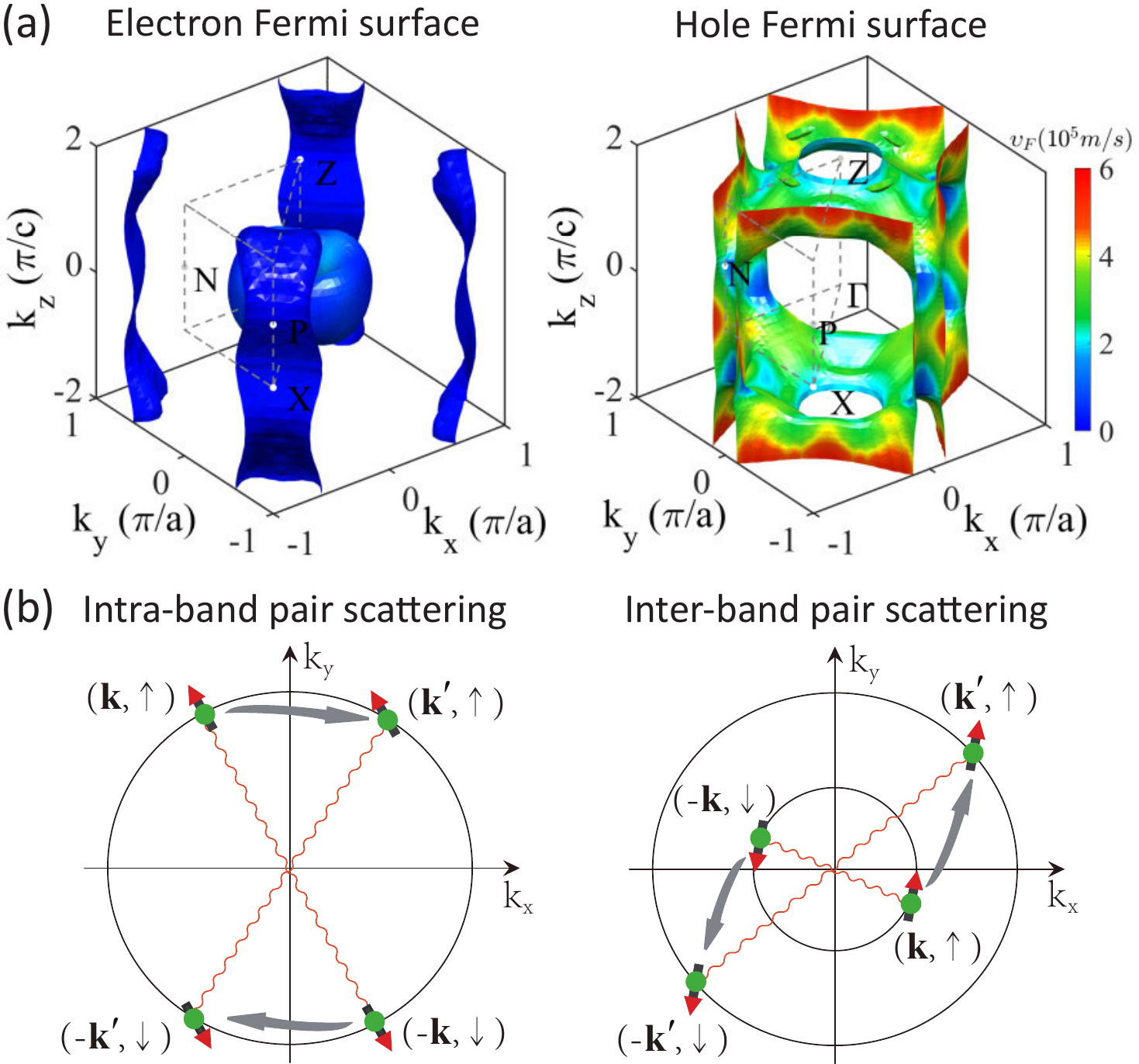}
\caption{(color online) (a) Fermi velocity distributions on the electron and hole Fermi surfaces of CeCu$_2$Si$_2$ at ambient pressure from DFT+$U$ calculations. The heavy electron Fermi surface contains mainly a corrugated-cylinder sheet near X point and a small torus. The hole Fermi surface is complex but also contains some $f$-orbital character. (b) Illustration of the intra- and interband scatterings (arrows) of the Cooper pairs. The wavy lines denote the electron pairs on the Fermi surfaces.}
\label{fig1}
\end{figure}

We start with the electronic band structure of CeCu$_2$Si$_2$ and calculate its Fermi surfaces using DFT+$U$ \cite{Perdew1996,Blaha2001,Anisimov1997}. The $f$-electrons are assumed to be itinerant and participate in the formation of the so-called "large" Fermi surfaces, in accordance with the experimental observation that superconductivity in CeCu$_2$Si$_2$ emerges near an itinerant spin density wave quantum critical point \cite{Sparn1998,Yuan2006,Stockert2011b}. The detailed band structures can be found in Ref.~\cite{Suppl}. The resulting Fermi surfaces are depicted in Fig.~\ref{fig1}(a) for $U=5\,$eV and contain two major parts: the heavy electron Fermi surface with a corrugated-cylinder sheet around the $X$ point (denoted by band 1) and a complex hole Fermi sheet (denoted by band 2), in agreement with previous calculations \cite{Zwicknagl1993,Ikeda2015}. We note that the hole Fermi surface is not completely light. It is partially hybridized and contains some portions of dominant $f$-characters. As discussed above, both parts of the Fermi surfaces are indispensable for a proper understanding of the superconductivity.

The fact that the two parts do not overlap allows us to ignore pairing between the two Fermi surfaces, since it involves a finite total moment. The electron pairs  on the two parts of Fermi surfaces are then connected by intra- and interband interactions \cite{Suhl1959,Dolgov2009}, as illustrated in Fig.~\ref{fig1}(b). We may write down the Eliashberg equations as \cite{Monthoux1992,Monthoux1999,Takimoto2004,Nishiyama2013}
\begin{eqnarray}
Z_{\mu}\left(\bold{k},i\omega_n \right)&=&1+\frac{\pi T}{\omega_n}\sum_{\nu,m} \int_{\text{FS}_\nu}
\frac{d\bold{k}^{\prime}_{\parallel}}{(2\pi)^3v_{\bold{k}^{\prime}_\text{F}}}\text{sgn} \left( \omega_m \right) \nonumber \\
&\times& V^{\mu\nu}\left( \bold{k}-\bold{k}^{\prime},i\omega_n-i\omega_m \right),
\label{eq1}\\
\lambda\phi_{\mu}\left(\bold{k},i\omega_n \right)&=&-\pi T\sum_{\nu,m}
\int_{\text{FS}_\nu}\frac{d\bold{k}^{\prime}_{\parallel}}{(2\pi)^3v_{\bold{k}^{\prime}_\text{F}}}\phi_{\nu} \left(\bold{k}^{\prime},i\omega_m \right) \nonumber \\
&\times& \frac{V^{\mu\nu}\left(\bold{k}-\bold{k}^{\prime},i\omega_n-i\omega_m \right)} {\left| \omega_m
Z_{\nu}\left(\bold{k}^{\prime},i\omega_m \right)\right|},
\label{eq2}
\end{eqnarray}
where $\mu$ and $\nu$ denote band indices, $\omega_{n/m}$ is the fermionic Matsubara frequency, the integral with FS$_\nu$ is over the Fermi surface of band $\nu$, $v_{\bold{k}^{\prime}_\text{F}}$ is the corresponding Fermi velocity, $Z_{\mu}$ is the renormalization function, and $\phi_{\mu}$ is the anomalous self-energy related to the superconducting gap function: $\Delta_{\mu}=\phi_{\mu}/Z_{\mu}$. $V^{\mu\nu}\propto\chi^{\mu\nu}$ represents the intra ($\mu=\nu$) or interband ($\mu\neq\nu$) pairing interactions, and $\chi^{\mu\nu}$ is the dynamical susceptibility due to quantum critical fluctuations. When solving the above equations, $\lambda$ can be viewed as the eigenvalue of the kernel matrix in the right-hand side of equation (\ref{eq2}). Each eigensolution represents a separate pairing channel and its eigenvector yields the corresponding gap structure on the Fermi surfaces \cite{Sigrist2005,Graser2009}. The leading pairing channel is determined by the largest eigenvalue $\lambda$ at a given $T_c\approx 0.6\,$K.

To proceed, we assume the following quantum critical form of the pairing interactions \cite{Millis1990}:
\begin{equation}
V^{\mu\nu}(\bold{q},i\nu_n)=\frac{V_0^{\mu\nu}}{1+\xi ^{2}(\bold{q}-\bold{Q}_\text{AF})^2 + |\nu_n|/\omega_{sf}},
\label{eq3}
\end{equation}
where $\nu_n$ is the bosonic Matsubara frequency. $V_0^{\mu\nu}$ are free parameters independent of $\bold{q}$ and $\nu_n$, which control the relative strength of the intra- and interband interactions. All other parameters can be determined from the experiment and we have the antiferromagnetic correlation length, $\xi\approx 25\,\rm{\mathring{A}}$, and the characteristic energy of spin fluctuations, $\omega _{sf}\approx 0.04\,$meV, which define an effective magnetic Fermi energy $\Gamma_{sf}=\omega_{sf}(\xi/a)^2\approx 1.5\,$meV $\gg T_c$ \cite{Stockert2011}. The pairing forces all peak at $\bold{Q}_\text{AF}$ in accordance with both experimental observation \cite{Stockert2011} and theoretical calculations \cite{Ikeda2015}.

\begin{figure}[t]
\centering\includegraphics[width=0.48\textwidth]{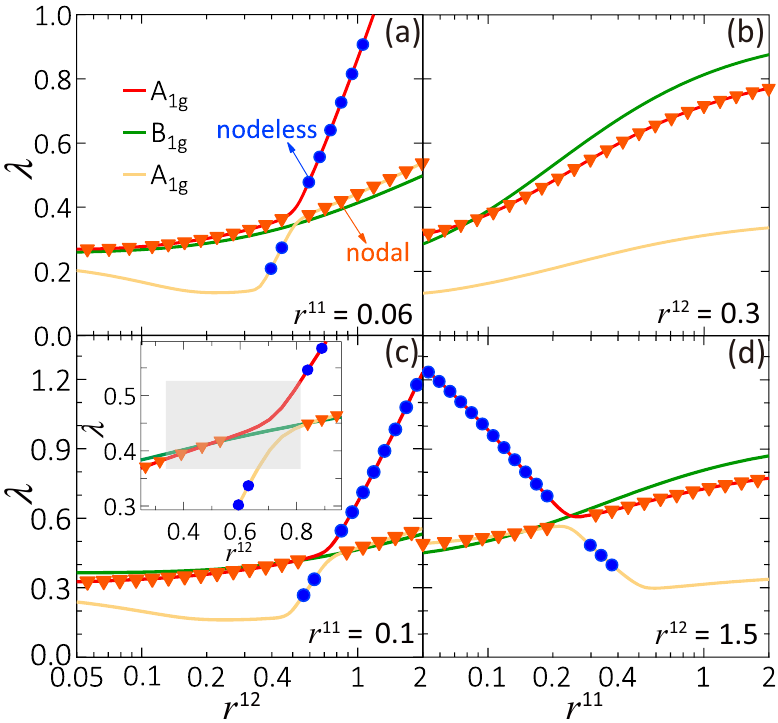}
\caption{(color online) Evolution of the eigenvalues $\lambda$ of three leading solutions as functions of $r^{11}$ or $r^{12}$. Two of the solutions belong to the same representation ($A_{1g}$) of the $D_{4h}$ group. Both are mixtures of nodeless ($\circ$) and nodal ($\triangledown$) $s$-waves. The other solution belongs to the $B_{1g}$ representation and has a $d_{x^2-y^2}$ symmetry. The inset in  (c) highlights the interplay of three solutions.}
\label{fig2}
\end{figure}

We solve the Eliashberg equations by approximating $\Delta(\bold{k},\omega_n)=\Delta(\bold{k},i\pi T_c)$ and using 2048 Matsubara frequencies and $47\times$47$\times$47 $\bold{k}$-meshes in the whole Brillouin zone. Figure~\ref{fig2} plots the eigenvalues of three leading pairing channels as a function of $r^{12}=V_0^{12}/V_0^{22}$ and $r^{11}=V_0^{11}/V_0^{22}$. The results are qualitatively similar regardless of the exact value of $V_0^{22}$. Among the three leading solutions, two are $s$-wave and belong to the same  representation of the $D_{4h}$ symmetry group ($A_{1g}$: 1, $k_x^2+k_y^2$, $k_z^2$, ...); one belongs to the $B_{1g}$ representation and is of $d_{x^2-y^2}$-wave. There exists an interesting interplay of these solutions with varying intra- and interband interactions. In Figs.~\ref{fig2}(a) and \ref{fig2}(c), the leading solution always increases with increasing $r^{12}$ for fixed $r^{11}$, indicating that the superconducting instability is always enhanced by the interband interaction. In contrast, for fixed $r^{12}=1.5$ and with increasing $r^{11}$, as shown in Fig.~\ref{fig2}(d), the leading $A_{1g}$ solution is first suppressed before it is taken over by the $B_{1g}$ solution. The transitions between different pairing channels occur in a narrow parameter range which is enlarged in the inset in Fig.~\ref{fig2}(c). While the $A_{1g}$ and $B_{1g}$ solutions belong to different representations and simply cross each other, the two $A_{1g}$ solutions belong to the same representation and are actually mixtures of nodal and nodeless $s$-wave solutions.

\begin{figure}[t]
\centering\includegraphics[width=0.48\textwidth]{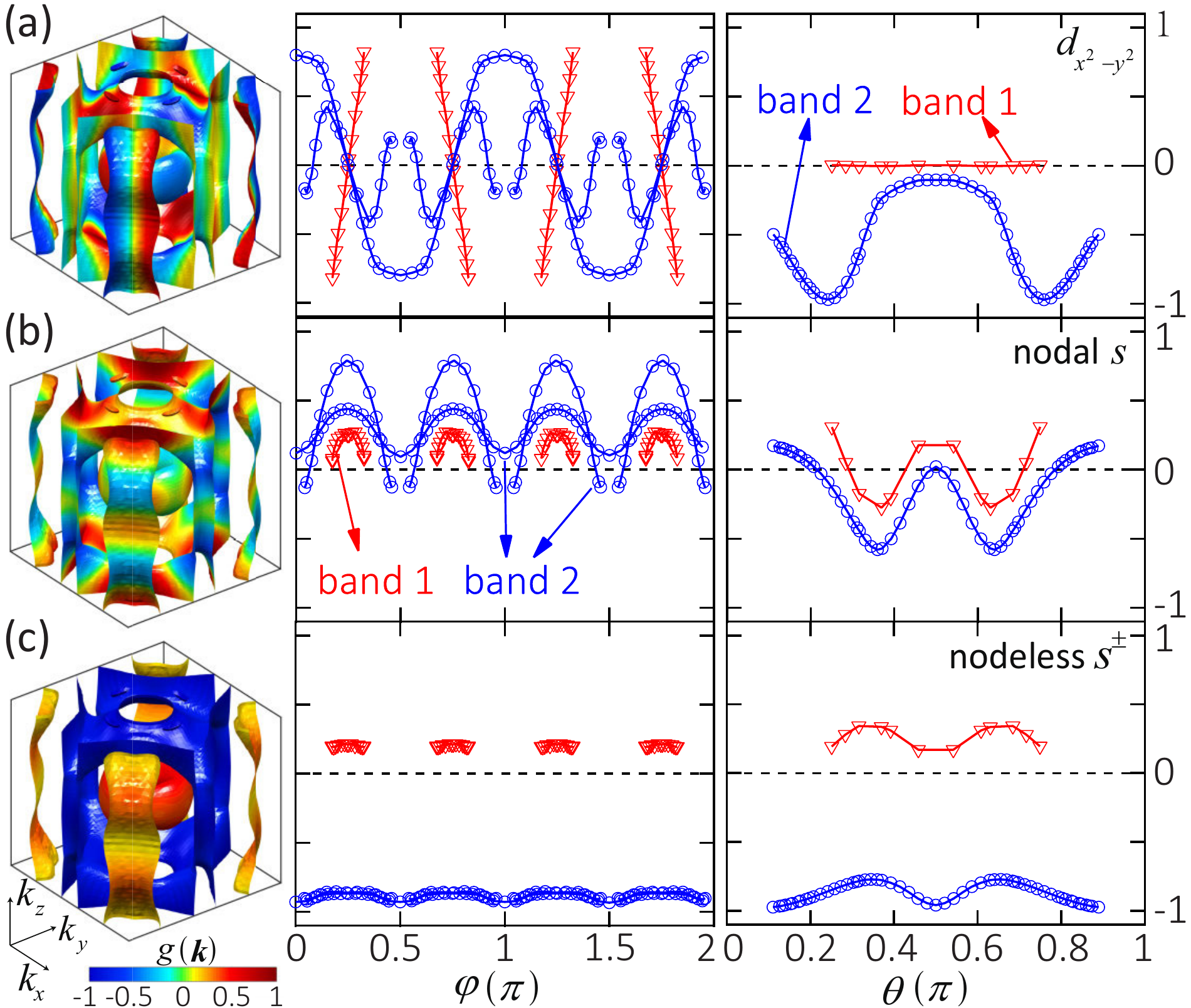}
\caption{(color online) Illustration of typical gap structures on the Fermi surfaces for fixed $r^{11}=0.1$: (a) $d_{x^2-y^2}$ at $r^{12}=0.3$; (b) nodal $s$ at $r^{12}=0.6$; (c) nodeless $s^{\pm}$ at $r^{12}=1.0$. The color bar represents the value of the eigenvector $g(\bold{k})$ normalized by its maximal value. For azimuthal ($\varphi$) dependence, we choose the $k_z=1.64\pi/c$ plane; while for polar ($\theta$) dependence, we choose the diagonal cut ($k_x=k_y$) for the heavy electron Fermi surface ($\triangledown$) and the $k_x=0$ plane cut for the hole Fermi surface ($\circ$). The torus is neglected for clarity.}
\label{fig3}
\end{figure}

To see this more clearly, we study in detail the gap structures of the leading solution for fixed $r^{11}=0.1$ as in Fig.~\ref{fig2}(c). Figure~\ref{fig3} plots the gap distributions on the whole Fermi surfaces. For small $r^{12}=0.3$, the $B_{1g}$ solution in Fig.~\ref{fig3}(a) shows an evident $d_{x^2-y^2}$-wave angular dependence on the azimuthal angle ($\varphi$). In particular, it contains line nodes along the vertical ($\theta$) direction for $k_x=k_y$. This is consistent with previous calculations for a single heavy electron band with nested Fermi surfaces \cite{Eremin2008}. Increasing $r^{12}$ leads to a phase transition to the $A_{1g}$ solution. For $r^{12}=0.6$ shown in Fig.~\ref{fig3}(b), we obtain an extended $s$-wave with nodal areas on both Fermi surfaces, similar to the loop-nodal $s^{\pm}$-wave solution derived in previous perturbative calculations \cite{Ikeda2015}. With further increasing $r^{12}$ and away from the avoided intersection as shown in the inset in Fig.~\ref{fig2}(c), the nodal areas gradually diminish as the nodal component becomes less important in the leading $A_{1g}$ solution. As shown in Fig.~\ref{fig3}(c) for $r^{12}=1.0$, the gap functions become almost angle-independent but with opposite signs on the heavy electron and hole Fermi surfaces, pointing to a nodeless $s^\pm$-wave symmetry. Moreover, the gap ratio is in reasonable agreement with experimental estimates \cite{Kittaka2014,Enayat2016,Pang2016,Kittaka2016}.

\begin{figure}[t]
\centering\includegraphics[width=0.48\textwidth]{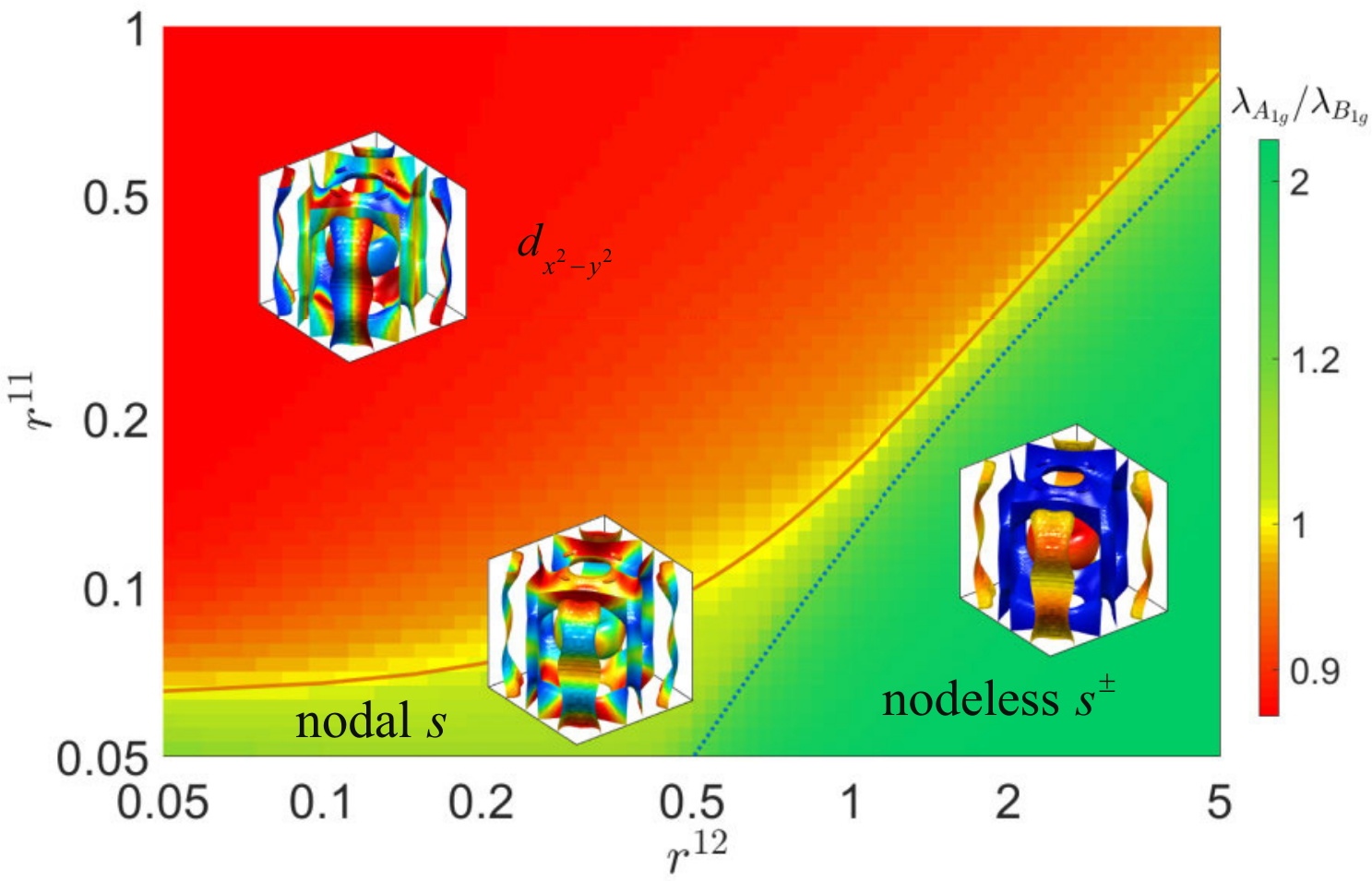}
\caption{(color online) A theoretical phase diagram for the superconductivity in CeCu$_2$Si$_2$ at ambient pressure. The axes denote the relative strengths of the intra- and interband interactions. The colorbar represents the ratio between the eigenvalues of the leading $A_{1g}$ and $B_{1g}$ solutions. The phase diagram is divided into three regions: $d_{x^2-y^2}$, nodal $s$ and nodeless $s^\pm$, illustrated with typical gap structures from Fig.~\ref{fig3}. The solid line marks the transition between $d_{x^2-y^2}$ and nodal $s$ and the dashed line shows the crossover from nodal $s$ to nodeless $s^\pm$.}
\label{fig4}
\end{figure}

Figure \ref{fig4} summarizes the above results in a global phase diagram of the superconductivity. We see two major regions with either $d_{x^2-y^2}$ or nodeless $s^\pm$-wave symmetry. In between, there exists a narrow region with a nodal $s$-wave solution. Obviously, for small $r^{12}$ and large $r^{11}$, superconducting pairing on the nested heavy electron Fermi surface dominates and $d_{x^2-y^2}$-wave is favored. This is, however, taken over by the nodeless $s^\pm$-wave for large $r^{12}$ when the interband interaction becomes strong. This nodeless structure is not accidental but required by the strong interband interaction to minimize the total Coulomb energy \cite{Mazin2009,Chubukov2015}. The nodal $s$-wave is prominent only for small $r^{11}$ and $r^{12}$, indicating that it is induced primarily by the hole Fermi surface. We remark that the pairing interaction between different Fermi surfaces is usually ignored in  simplified considerations of heavy fermion superconductivity. In comparison with previous theories \cite{Eremin2008,Ikeda2015}, our results suggest that it could be important in realistic multiband systems where orbital or valence fluctuations could become important and cause significant interband pairing interactions between different parts of the hybridized Fermi surfaces \cite{Varma2012}. This might be relevant in CeCu$_2$Si$_2$ \cite{Jaccard1999,Holmes2004,Miyake2014} and PuCoGa$_5$ \cite{Ramshaw2015,Bauer2012}, in contrast to the case of CeCoIn$_5$ and CeRhIn$_5$ \cite{Allan2013,Yang2014,Yang2015}. 

Such a difference is largely associated with the special orbital property of the hole Fermi surface in CeCu$_2$Si$_2$. Different from the general understanding since early days, both the electron and hole Fermi surfaces belong to hybridization bands and contain significant $f$-orbital character, thus allowing for strong interband scattering from the $f$-orbital interactions when mapped to the band basis. While calculations based on DFT+$U$ already yield strong interband scattering, as manifested by the nodal $s$-wave solution in Ref.~\cite{Ikeda2015} and our phase diagram, more sophisticated dynamical mean-field theory calculations have predicted a much weaker nesting effect of the electron Fermi surface and in the mean time a strong interplay between the $f$-orbitals governing the electron and hole Fermi surfaces \cite{Pourovskii2014}, which further disfavors the $d$-wave solution. Combining these may provide key information for parametrizing the effective intra- and interband interactions and thus help to derive a better microscopic understanding of the superconductivity in CeCu$_2$Si$_2$.

Our derived $s^{\pm}$-wave solution provides a plausible theoretical basis for the observation of two nodeless superconducting gaps in CeCu$_2$Si$_2$. The opposite sign on the heavy electron and hole Fermi surfaces explains the neutron spin resonance mode below $T_c$ \cite{Stockert2011}. On the other hand, the NMR spin-lattice relaxation rate exhibits no clear coherence peak \cite{Kitaoka1986,Fujiwara2008}, which has often been used to argue against the $s^\pm$-wave \cite{Pang2016}. However, the coherence peak may be suppressed in the presence of a multiband effect \cite{Bang2017}. As a result, $s^\pm$-wave can actually yield a good fit to the NMR data as discussed in a recent experimental analysis \cite{Kitagawa2017}.

An alternative 'd+d'-wave has recently been proposed following the study of iron-pnictide superconductors \cite{Pang2016,Nica2017}. Its gap structures contain simultaneously an intraband $d_{x^2-y^2}$-wave component and an interband $d_{xy}$-wave component. In our calculations, the interband component involves finite-moment pairing and is insignificant owing to the lack of Fermi surface overlap. To see this, we write down the linearized gap equation for the interband pairing function: $\Phi_{\bold{k}}=-\sum_{\bold{k'}}W_{\bold{k},\bold{k'}}^{12}\Phi_{\bold{k'}} (\text{tanh}\frac{\xi_{\bold{k'}}^+}{2T}+\text{tanh}\frac{\xi_{\bold{k'}}^-}{2T})/(\xi_{\bold{k'}}^{+}+\xi_{\bold{k'}}^{-})$, in which $W_{\bold{k},\bold{k'}}^{12}$ is the corresponding pairing interaction, $\xi_{\bold{k'}}^{\pm}=\frac{1}{2}|\epsilon_{\bold{k'}1}+\epsilon_{\bold{k'}2}|\pm\frac{1}{2}(\epsilon_{\bold{k'}1}-\epsilon_{\bold{k'}2})$, and $\epsilon_{\bold{k'}\mu}$ is the quasiparticle dispersion. Our band calculations give a minimal energy difference: $\delta=\min_{\bold{k'}} |\epsilon_{\bold{k'}1}-\epsilon_{\bold{k'}2}| \approx 60\,$meV, which is much larger than $T_c$ ($0.6\,$K) and the characteristic magnetic energy $\Gamma_{sf}$ ($1.5\,$meV) in CeCu$_2$Si$_2$ at ambient pressure. Hence the right-hand side $(\text{tanh}\frac{\xi_{\bold{k'}}^+}{2T}+\text{tanh}\frac{\xi_{\bold{k'}}^-}{2T})/(\xi_{\bold{k'}}^{+}+\xi_{\bold{k'}}^{-})\sim 1/\delta$ near the Fermi surfaces and exhibits no divergence or Cooper instability as $T\rightarrow 0$ \cite{Fischer2013}. An $s^{++}$-wave has also been proposed based on recent electron irradiation experiments in which $T_c$ was found to be suppressed only weakly by Ce defects \cite{Yamashita2017}. Whether or not this can be reconciled with earlier observations showing significant $T_c$-suppression by nonmagnetic impurities remains unclear \cite{Spille1983,Adrian1987,Yuan2004}. In our simple model, a nodeless $s^{++}$-wave may also be obtained at large negative $r^{12}$. However, this solution is unstable and leads to a negative renormalization function.

In conclusion, we propose a phenomenological model to study the superconducting gap symmetry of CeCu$_2$Si$_2$ at ambient pressure and succeed in obtaining a nodeless $s^\pm$-wave solution for a strong interband interaction. Our predicted gap ratio agrees well with the experiment. Contrary to prevailing ideas, we find that the $s$-wave solution is primarily associated with the hole Fermi surface. Thus at high pressures, where only the hole Fermi surface exists, a nodal $s$-wave solution is expected for the second superconducting dome of CeCu$_2$Si$_2$. With reduced pressure, the electron Fermi surface emerges, causing significant interband scattering and, consequently, a transition to the nodeless $s^\pm$-wave superconductivity. This is a different picture from the conventional one and provides a possible basis for solving the current discrepancy between the theory and experiment. It highlights the potential importance of multiple Fermi surfaces and strong interband scattering which may cause rather peculiar and unexpected behaviors in heavy fermion superconductors.

This work was supported by the National Key R\&D Program of China (Grant No. 2017YFA0303103), the National Natural Science Foundation of China (NSFC Grant No. 11774401, No. 11522435, No. 11674025, No. 11274041, and No. 11334012), the State Key Development Program for Basic Research of China (Grant No. 2015CB921303), the Strategic Priority Research Program (B) of the Chinese Academy of Sciences (CAS) (Grant No. XDB07020200) and the Youth Innovation Promotion Association of CAS.

\clearpage
\includegraphics[trim=2.4cm 0 0 0]{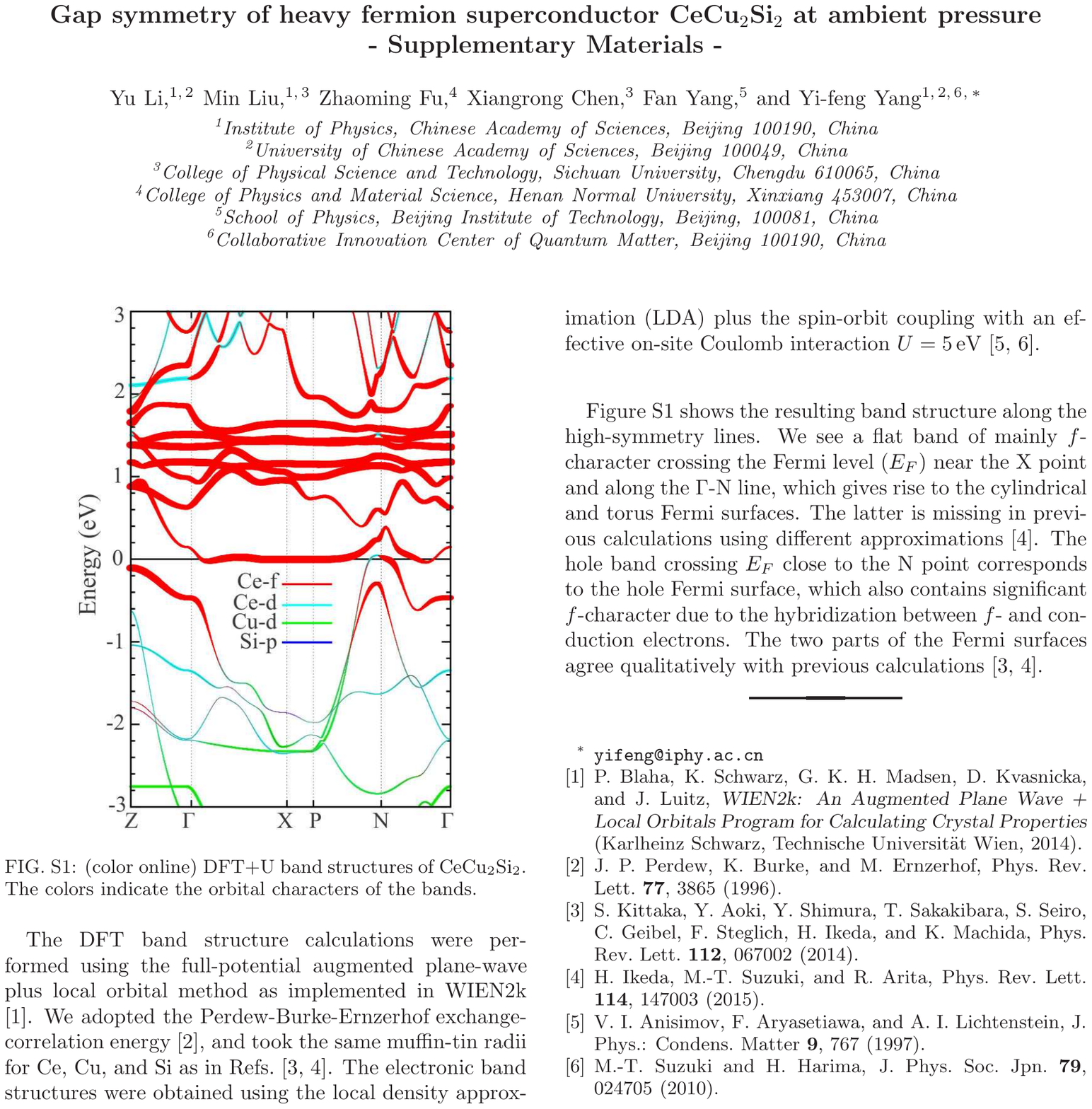}

\end{document}